\documentclass[12pt]{article}

\usepackage{amsmath}
\usepackage{amssymb}
\usepackage{amsthm}
\usepackage{bm}
\usepackage{graphicx}

\theoremstyle{definition}
\newtheorem{theorem}{Theorem}
\newtheorem{lemma}{Lemma}

\title{Extension of the Lieb-Schultz-Mattis theorem}
\author{Kiyohide Nomura, Junpei Morishige and Takaichi Isoyama
\\
Department of Physics, Kyushu University
\\
Fukuoka 812-8581,
JAPAN
\\
E-mail:knomura@stat.phys.kyushu-u.ac.jp
}
\date{}
\begin{document}
\maketitle

\begin{abstract}
 
Lieb, Schultz and Mattis (LSM)
\cite{Lieb-Schultz-Mattis-1961}
studied the S=1/2 XXZ spin chain.
The theorems of LSM's paper can be applied to broader models.
In the original LSM theorem the nonfrustrating system was assumed. 
However, reconsidering the LSM theorem, we can extend the LSM theorem
 for frustrating systems.
Next, several researchers tried to extend the LSM theorem
for excited states.
In the cases $S^{z}_{T} = \pm 1,\pm 2 \cdots$, the lowest energy
eigenvalues are continuous for wave number $q$.
But we found that their proofs were insufficient, and improve upon them.
In addition, we can prove the LSM theory without the assumption of the
 discrete symmetry, which means that the LSM-type theorems
 are applicable for 
 Dzyaloshinskii-Moriya type interactions or other nonsymmetric models.
 
 \medskip
 Keywords: Lieb-Schultz-Mattis, rigorous theorem, frustration,
 one-dimension, Dzyalosinskii-Moriya
\end{abstract}

\section{Introduction}

In statistical physics, exact solutions such as Onsager's
theory for a two-dimensional (2D) Ising model,
the Bethe ansatz for 1D quantum
systems \cite{Bethe-1931,des-Cloizeaux-Pearson-1962,Yang-Yang-1966,Yamada-1969},
the matrix product method
\cite{Majumdar-Ghosh-1969,Affleck-Kennedy-Lieb-Tasaki-1988},
etc, have played important roles.

Besides exact solutions, there are rigorous theorems
 such as the Mermin-Wagner theorem, the Marshall-Lieb-Mattis theorem
\cite{Marshall-1955,Lieb-Mattis-1962},
 the Lieb-Schultz-Mattis theorem \cite{Lieb-Schultz-Mattis-1961}
 etc,
 which do not give quantitative, but qualitative results.
 Since such rigorous theorems are based on symmetries,
 they can be applied to broader models.
 And one can use them to check the consistency of approximations,
 experiments, or numerical data.

 Lieb, Schultz and Mattis (LSM)
\cite{Lieb-Schultz-Mattis-1961}
studied the S=1/2 XXZ spin chain.
In appendix B of their paper, they described two theorems.
In the first theorem, it was proved that the ground state is unique
in finite $L$ systems.
In the second theorem, they proved
that there exists a low-energy $O(1/L)$
 excited state;
 in the infinite limit,
this means that either there are degenerate ground states or a vanishing gap.

The first theorem was nothing more than an extension of Marshall's theorem
\cite{Marshall-1955}, and was later applied to more general cases
(higher dimensions, bipartite lattice, ferrimagnetism, etc.)
by \cite{Lieb-Mattis-1962},
therefore it is appropriate to call the first theorem
the ``Marshal-Lieb-Mattis (MLM) theorem''.
The MLM theorem was applied to the spin model with biquadratic exchange \cite{Munro-1976}.
Since the Perron-Frobenius theorem is used in the MLM theorem,
it is limited in the non-frustrated case.

The second LSM theorem was extended for general spin $S$ 
and was applied for various models by \cite{Affleck-Lieb-1986},
and  it was proven that there exists a low-energy $O(1/L)$ excited state
for half-odd-integer spin cases.
It is worth noting that the assumptions of the MLM theorem
and those of the second LSM theorem are independent,
though this fact has been overlooked. 
Hereafter we call the second LSM theorem alone the LSM theorem.
In this paper we shall extend the LSM theorem 
without the assumption of the uniqueness of the ground state,
using a squeeze theorem type discussion.
Therefore, we can extend the LSM theorem for frustrated systems.

On the other hand, 
Kolb \cite{Kolb-1985} studied the energy spectra of the XXZ spin
chain with the twist boundary condition, independent of
\cite{Lieb-Schultz-Mattis-1961,Affleck-Lieb-1986}. 
He obtained similar conclusions regarding the two-fold degenerate ground
state for the $S$ half-odd-integer case.
He also pointed out the change of the wave vector when varying the twist
boundary condition for nonzero total spin $S^{z}_{T}$ cases.
In section II of \cite{Fath-Solyom-1993},
combining Kolb's idea and LSM theorem,
F\'ath and S\'olyom
insisted that they proved the continuity of the dispersion curve
for $S^{z}_{T} = \pm 1,\pm 2 \cdots$. 
However, we find that their proof is insufficient.

Oshikawa {\it et al.}
\cite{Oshikawa-Yamanaka-Affleck-1997},
using the LSM type discussion,
pointed out that there may be magnetic plateaux. 
In addition, they emphasized the importance of either the space inversion or the spin
reversal symmetry, besides the U(1) and the translational symmetry.
Although the proof of the LSM theorem becomes simplified
with the assumption of discrete symmetry,
it excludes nonsymmetric spin models.

In this paper we prove the LSM theorem,
by using only the assumptions of the U(1) symmetry,
the translational invariance and the short-range interaction.
There is no need for the uniqueness of the ground state for a finite system,
the space inversion or the spin reversal symmetry assumptions,
which means we can extend the LSM theorem for
frustrating or nonsymmetric models
(e.g. Dzyaloshinskii-Moriya interaction \cite{Dzyaloshinskii-1958,Moriya-1960}).

The layout of the paper as follows.
In section 2, we introduce the definition of symmetry operations.
Section 3 is the main part of this work:
we prove the continuity and the periodicity of the lowest energy spectra 
as a function of wave number $q$,
assuming the U(1) and the translational symmetries
plus the short-range interaction.
Furthermore, in section 4 we discuss the discrete symmetries,
i.e. the space inversion and  the spin reversal.
We consider several specific spin models with various lattice structures and
symmetries in section 5.
In section 6, using the nonfrustration condition (model specific nature),
we can discuss the minimum location of spectra. 
In section 7, we compare our theorems with
\cite{Lieb-Schultz-Mattis-1961,Affleck-Lieb-1986} and
\cite{Oshikawa-Yamanaka-Affleck-1997}.
In section 8, we illustrate possible spectra for frustrating systems,
those for magnetization plateau, and those for nonsymmetric case.
Section 9 is the conclusion.

\section{Model, symmetries, eigenstates}

In this section we consider the symmetries of the spin chain.
As a typical model, we treat the following 
1D generalized XXZ spin Hamiltonian:
\begin{align}
 \hat{H} &= \sum_{j=1}^{L} \sum_{r=1}^{L/2}
\left( \frac{J(r)}{2}(\hat{S}^{+}_{j}
  \hat{S}^{-}_{j+r}+\hat{S}^{-}_{j} \hat{S}^{+}_{j+r}  )
  + \Delta(r)  \hat{S}^{z}_{j} \hat{S}^{z}_{j+r}
\right)+ h\sum_{j=1}^{L} \hat{S}^{z}_{j},
\label{eq:Hamiltonian-generalized-XXZ}
\end{align}
where
$
 (\hat{\bm{S}}_{j})^{2}= S(S+1)
$
($S=1/2,1,\cdots$),
with the system size $L$ even
and the
periodic boundary condition (PBC):
\begin{equation}
 \hat{S}^{x,y,z}_{L+j} =\hat{S}^{x,y,z}_{j}.
\end{equation}
Hamiltonian (\ref{eq:Hamiltonian-generalized-XXZ})
is invariant under the space
inversion.

And when $h=0$ in (\ref{eq:Hamiltonian-generalized-XXZ}),
the Hamiltonian is invariant under the spin reversal.
       
\subsection{Symmetries}

Next we enumerate the symmetry operations.
Hereafter we denote
\begin{equation}
\hat{S}^{x}_{T} \equiv \sum_{j=1}^{L} \hat{S}^{x}_{j},
\quad
\hat{S}^{y}_{T} \equiv \sum_{j=1}^{L} \hat{S}^{y}_{j},
\quad
\hat{S}^{z}_{T} \equiv \sum_{j=1}^{L} \hat{S}^{z}_{j}.
\end{equation}

\begin{enumerate}
 \item{Rotation around the $z$-axis}

We define the rotational operator of the $z$-axis as   
 \begin{align}
\hat{U}^{z}_{\theta} & \equiv  \exp(- i \theta   \hat{S}^{z}_{T}),
\\
  (\hat{U}^{z}_{\theta})^{\dagger} \hat{S}^{\pm}_{j} \hat{U}^{z}_{\theta}
  &=
\hat{S}^{\pm}_{j} \exp(\pm i\theta),
\quad
(\hat{U}^{z}_{\theta})^{\dagger} \hat{S}^{z}_{j} \hat{U}^{z}_{\theta} =  \hat{S}^{z}_{j}.
\end{align}

\item{Translation operator by one-site: $\hat{U}_{\rm trl}$.} 

\begin{equation}
 \hat{U}_{\rm trl}^{\dagger} \hat{S}^{x,y,z}_{j}  \hat{U}_{\rm trl} = 
 \hat{S}^{x,y,z}_{j+1}.
\end{equation}

 \item{Space inversion (site parity).} 
\begin{equation}
 \hat{P}^{\dagger} \hat{S}^{x,y,z}_{j}  \hat{P} = \hat{S}^{x,y,z}_{L-j}.
\end{equation}
Because $\hat{P}^{2}=1$,
it is shown that 
$\hat{P}^{-1} = \hat{P}^{\dagger}=\hat{P}$ 
and the eigenvalue of
$\hat{P}$
is
$\pm 1$.

\item {Space inversion (link parity).}
\begin{equation}
 \hat{P}_{\rm link}^{\dagger}\hat{S}^{x,y,z}_{j}  \hat{P}_{\rm link} = \hat{S}^{x,y,z}_{L-j+1}.
\end{equation}
The link parity operator can be defined as the product of the site
parity operator
and the translational operator.
\begin{equation}
  \hat{P}_{\rm link}= \hat{P} \hat{U}_{\rm trl}. 
\end{equation}

\item{Relation between site parity and translation.} 

There is a relation between parity and the translation operation:
\begin{equation}
 \hat{P}\hat{U}_{\rm trl} \hat{P} = \hat{U}_{\rm trl}^{-1},
\label{eq:Translation-Parity-Relation}
     \end{equation}
therefore
\begin{equation}
 ( \hat{P} \hat{U}_{\rm trl})^{2} =1,
\end{equation}
that is, the eigenvalue of link parity is also $\pm 1$.

 \item{Spin reversal} 

The operator of $\pi$ rotation around the $y$-axis is given as
\begin{equation}
\hat{U}^{y}_{\pi}  \equiv  \exp(-\pi i  \hat{S}^{y}_{T})
\end{equation}
then
\begin{equation}
(\hat{U}^{y}_{\pi})^{\dagger} \hat{S}^{\pm}_{j} \hat{U}^{y}_{\pi} =
 -\hat{S}^{\mp}_{j},
\quad
(\hat{U}^{y}_{\pi})^{\dagger} \hat{S}^{z}_{j} \hat{U}^{y}_{\pi} =
-\hat{S}^{z}_{j}.
\label{eq:y-axis-pi-inversion}
\end{equation}

The eigenvalue of the operator $\hat{S}^{y}_{T}$ is an integer from the
evenness of $L$.
Therefore,  we obtain that $(\hat{U}^{y}_{\pi})^{2} =1$ and
$(\hat{U}^{y}_{\pi})^{-1} = (\hat{U}^{y}_{\pi})^{\dagger} = \hat{U}^{y}_{\pi}$
and that the eigenvalue of
$\hat{U}^{y}_{\pi}$ is $\pm 1$.

 \item

The operators 
$\hat{S}^{x,y,z}_{T} $
are invariant under the translation and the space inversion.    
In addition,
$\hat{S}^{z}_{T} $
is invariant under the rotation around the $z$ axis.

\end{enumerate}

\subsection{Vacuum}

We take the fully aligned spin state as a vacuum:
\begin{align}
 \hat{S}^{z}_{j} | 0 \rangle = S  | 0 \rangle,
\quad
 \hat{S}^{+}_{j} | 0 \rangle = 0,
\quad
 \langle 0 | 0 \rangle =1.
 \label{eq:Vacuum}
\end{align}

\subsection{Eigenstates}
We write the eigenstate for the translation and the total spin $\hat{S}^{z}_{T}$
as
\begin{align}
\hat{S}^{z}_{T}|S^{z}_{T};  q\rangle = S^{z}_{T} |S^{z}_{T};  q\rangle,
\quad
 \hat{U}_{\rm trl}|S^{z}_{T}; q \rangle = \exp(iq) |S^{z}_{T};  q\rangle.
\end{align}
Moreover, when the Hamiltonian is translational and $U(1)$
invariant,
one can write
\begin{align}
 \hat{H} |S^{z}_{T};  q\rangle &= E(S^{z}_{T};  q) |S^{z}_{T};  q\rangle
\end{align}

\subsection{Theorem on the translation, space inversion and rotation
  around the $y$ axis in PBC}

We review several theorems of the PBC case.
\begin{theorem}

\begin{enumerate}
 \item
      Energy spectra are $2\pi$ periodic with wave number $q$.
      \begin{equation}
       E(S^{z}_{T};  q+2\pi) = E(S^{z}_{T};  q).
      \end{equation}

 \item 
       For the wave number $q\neq 0,\pi$,
     by using (\ref{eq:Translation-Parity-Relation}),
       one can show
       \begin{equation}
	\hat{P} | S^{z}_{T};  q \rangle =  | S^{z}_{T};  -q \rangle.
       \end{equation}

Parity of the eigenstate $q=0$ or $q=\pi$ is well defined:
\begin{align}
 \hat{P}|S^{z}_{T};  q=0 \rangle = \pm |S^{z}_{T};  q=0 \rangle,
\notag \\
 \hat{P}|S^{z}_{T};  q=\pi \rangle = \pm |S^{z}_{T};  q=\pi \rangle.
\end{align}

 \item
      For $S^{z}_{T} \neq 0$,
     by using (\ref{eq:y-axis-pi-inversion}), one can show
\begin{equation}
 \hat{U}^{y}_{\pi} | S^{z}_{T};  q \rangle =  | - S^{z}_{T};  q \rangle.
\end{equation}

The eigenstate of $S^{z}_{T} = 0$ is also the eigenstate of
       $\hat{U}^{y}_{\pi} $ with eigenvalue $\pm 1$.
\begin{equation}
 \hat{U}^{y}_{\pi} |S^{z}_{T}=0;  q \rangle =  \pm |S^{z}_{T}=0;  q \rangle.
\end{equation}

\item
\begin{equation}
 (\hat{P}\hat{U}^{y}_{\pi})^2 
= \hat{P}\hat{U}^{y}_{\pi}\hat{P}\hat{U}^{y}_{\pi}
= \hat{U}^{y}_{\pi}\hat{U}^{y}_{\pi}
= 1
\end{equation}
     thus the eigenvalue of $\hat{P}\hat{U}^{y}_{\pi}$ is $\pm 1$.

 \item When the Hamiltonian is invariant under the space inversion,
       
       energy spectra are symmetric about $q=0$: 
       \begin{equation}
	E(S^{z}_{T};  -q) = E(S^{z}_{T};  q).
       \end{equation}
       
 \item When the Hamiltonian is invariant under the spin reversal,
       
energy spectra are symmetric under the spin reversal:
\begin{equation}
 E(-S^{z}_{T};  q) = E(S^{z}_{T};  q).
\end{equation}

\end{enumerate}
\end{theorem}

\section{Extension of the LSM theorem}

In this section, we will extend the LSM theorem
without the assumption of the uniqueness of the lowest energy,
by using squeeze theorem type methods.
And we will use only the translational
and the U(1) symmetry, and will not treat the Hamiltonian directly.
We do not assume the discrete symmetries such as
the space inversion or the spin reversal.
Hereafter we express $|S^{z}_{T}; q\rangle$
as one of the lowest energy eigenstates in the subspace of $S^{z}_{T}$ and $q$.

We define the twisting unitary operator as
\begin{equation}
 \hat{U}^{\rm tw}_{\pm 2\pi} \equiv \exp \left(\mp \frac{2\pi i}{L} \sum_{j=1}^{L} j (\hat{S}^{z}_{j}
				      -S)\right),
\label{eq:Twisting-Operator}
\end{equation}
then we obtain
\begin{align}
(\hat{U}^{\rm tw}_{\pm 2\pi})^{\dagger} \hat{S}^{+}_j \hat{U}^{\rm tw}_{\pm 2\pi}
&=  \hat{S}^{+}_j \exp(\pm 2 \pi i j /L),
\notag \\
(\hat{U}^{\rm tw}_{\pm 2\pi})^{\dagger} \hat{S}^{-}_j \hat{U}^{\rm tw}_{\pm 2\pi}
&=  \hat{S}^{-}_j \exp(\mp 2 \pi i j /L),
\notag \\
(\hat{U}^{\rm tw}_{\pm 2\pi})^{\dagger} \hat{S}^{z}_j \hat{U}^{\rm tw}_{\pm 2\pi}
&=  \hat{S}^{z}_j, 
\end{align}
and
\begin{equation}
 (\hat{U}^{\rm tw}_{2\pi})^{-1} =(\hat{U}^{\rm tw}_{2\pi})^{\dagger} =\hat{U}^{\rm tw}_{-2\pi},
\end{equation}
and
\begin{equation}
 \hat{U}^{\rm tw}_{2\pi} |0 \rangle = |0 \rangle.
\end{equation}

Doing unitary transform (\ref{eq:Hamiltonian-generalized-XXZ}) with twisting
operator,
we obtain
\begin{align}
 &(\hat{U}^{\rm tw}_{\pm 2\pi})^{\dagger} \hat{H}  \hat{U}^{\rm tw}_{\pm 2\pi}
 -\hat{H}
\notag \\
 &= \sum_{j=1}^{L} \sum_{r=1}^{L/2} 
 \frac{J(r)}{2}
 (\hat{S}^{+}_{j} \hat{S}^{-}_{j+r} (\exp(\mp 2\pi r i/L)-1)
+ h.c.).
 \label{eq:Hamiltonian-TBC}
\end{align}

\subsection{Main theorem}

\begin{lemma}
 (Translation operator and twisting operator)

\begin{align}
\hat{U}^{\rm tw}_{\pm 2\pi} \hat{U}_{\rm trl}
&=  \hat{U}_{\rm trl}\hat{U}^{\rm tw}_{\pm 2\pi}
 \exp\left(\pm \frac{2 \pi i}{L}(\hat{S}^{z}_{T}
 -SL)\right),
\notag \\
 \hat{U}_{\rm trl} \hat{U}^{\rm tw}_{\pm 2\pi}
&=   \hat{U}^{\rm tw}_{\pm 2\pi} \hat{U}_{\rm trl}
 \exp\left(\mp \frac{2 \pi i}{L}(\hat{S}^{z}_{T}
 -SL)\right).
\label{eq:TBC-translation-2pi}
\end{align}

\begin{proof}
 
\begin{align}
\hat{U}_{\rm trl}^{\dagger} \hat{U}^{\rm tw}_{\pm 2\pi} \hat{U}_{\rm trl}
&=\exp\left(\mp \frac{2\pi i}{L}\sum_{j=1}^{L} j(\hat{S}^{z}_{j+1} -S)\right)
\notag \\
&=\exp\left(\mp\frac{2\pi i}{L}\left(\sum_{j=2}^{L} (j-1)(\hat{S}^{z}_{j} -S)
+L(\hat{S}^{z}_{L+1}-S)\right)
\right)
\notag \\
&=\hat{U}^{\rm tw}_{\pm 2\pi} \exp\left(\pm \frac{2\pi i}{L}(\hat{S}^{z}_{T}
 -SL)\right)\exp(\mp 2\pi i(\hat{S}^{z}_{1} -S)),
\end{align}
 where we used  $\hat{S}^z_{L+1} = \hat{S}^z_{1} $.
 Combining this equation with the fact that
the eigenvalue of $\hat{S}^{z}_{1} -S$ is an integer,
we obtain
(\ref{eq:TBC-translation-2pi}).

\end{proof}
\end{lemma}

\begin{theorem}

In the subspace with a quantum number $S^{z}_{T}$,
on the lowest energies of the three wave numbers 
$q$, $q\pm 2\pi S^{z}_{T}/L +2\pi S$,
the following inequality holds:
\begin{equation}
E(S^{z}_{T};  q-2\pi S^{z}_{T}/L +2\pi S)
+E(S^{z}_{T} ;  q+2\pi S^{z}_{T}/L +2\pi S)
- 2 E(S^{z}_{T}; q)
\le O(1/L).
\label{eq:Energy-Spectrum-Inequality}
\end{equation}

\begin{proof}
 
 The following combination
\begin{align}
(\hat{U}^{\rm tw}_{2\pi})^{\dagger} \hat{H} \hat{U}^{\rm tw}_{2\pi}  
+(\hat{U}^{\rm tw}_{-2\pi})^{\dagger} \hat{H} \hat{U}^{\rm tw}_{-2\pi}  
 -2\hat{H}
 \label{eq:Double-Twisting-Hamiltonian}
\end{align}
is translational invariant from lemma 1.

  And from lemma 1, we obtain
 \begin{equation}
  \hat{U}_{\rm trl} (\hat{U}^{\rm tw}_{\pm 2\pi}|S^{z}_{T}; q\rangle)
   = \exp(i(q \mp 2\pi S^{z}_{T} /L + 2\pi S))
   (\hat{U}^{\rm tw}_{\pm 2\pi}|S^{z}_{T}; q\rangle)
   \label{eq:Wave-number-eigenstate-of-twisting}
 \end{equation}
since $2S$ is an integer.
 
Using (\ref{eq:Double-Twisting-Hamiltonian}) and
(\ref{eq:Wave-number-eigenstate-of-twisting}),
we can prove the following inequality:
\begin{align}
& E(S^{z}_{T}; q-2\pi S^{z}_{T}/L +2\pi S)
+E(S^{z}_{T}; q+2\pi S^{z}_{T}/L +2\pi S)
- 2 E(S^{z}_{T}; q)
\notag \\
& \le 
\langle  S^{z}_{T}; q  | 
((\hat{U}^{\rm tw}_{2\pi})^{\dagger} \hat{H} \hat{U}^{\rm tw}_{2\pi}  
+(\hat{U}^{\rm tw}_{-2\pi})^{\dagger} \hat{H} \hat{U}^{\rm tw}_{-2\pi}  
- 2 \hat{H} )
| S^{z}_{T}; q  \rangle
\notag \\
&
=
\sum_{j=1}^{L}\sum_{r=1}^{L/2} J(r)
(\cos(2\pi r/L)-1)
 \langle  S^{z}_{T}; q  |
(\hat{S}^{+}_{j}\hat{S}^{-}_{j+r} +\hat{S}^{-}_{j}\hat{S}^{+}_{j+r} )
|S^{z}_{T}; q  \rangle
\notag \\
&
\le O(1/L),
\label{eq:Energy-Spectrum-Inequality2}
\end{align}
 where we used the variational principle,
 and the next relation
 \begin{equation}
   \langle  S^{z}_{T}; q  |
(\hat{S}^{+}_{j}\hat{S}^{-}_{j+r} +\hat{S}^{-}_{j}\hat{S}^{+}_{j+r} )
|S^{z}_{T}; q  \rangle
=    \langle  S^{z}_{T}; q  |
(\hat{S}^{+}_{1}\hat{S}^{-}_{1+r} +\hat{S}^{-}_{1}\hat{S}^{+}_{1+r} )
|S^{z}_{T}; q  \rangle,
 \end{equation}
 using translational operations,
  and we assume that the transverse interaction is short-range
 (for example $|J(r)| \propto \exp (- m|r|)$.
More detailed discussion on the interaction range is found in
\cite{Hakobyan-2003}.
\end{proof}
\end{theorem}

[Remark]
\begin{itemize}
 \item 
The longitudinal interaction $\Delta(r)$ and the magnetic field $h$
give no restriction on theorem 2.
\item
Although the form in line 3 of
(\ref{eq:Energy-Spectrum-Inequality2})
seems specific for the model (\ref{eq:Hamiltonian-generalized-XXZ}),
one can show (\ref{eq:Energy-Spectrum-Inequality2}) for multibody interactions etc.
These interactions are expressed as a sum of terms
\begin{equation}
  \hat{S}^{+}_{j}  \hat{S}^{-}_{j+r_1}
   \hat{S}^{+}_{j+r_2}  \hat{S}^{-}_{j+r_3}\cdots
\end{equation}
where the number of the raising operators should be equal to the number of
the lowering operators from the U(1) symmetry.
Then it is easy to show the inequality 
(\ref{eq:Energy-Spectrum-Inequality2}).

\end{itemize}

\subsection{Continuity of energy spectra}

  \begin{theorem}
   
The lowest energy spectra in the subspace
of $S^{z}_{T}=(S-m/n)L+\Delta S^{z}_{T}$ \;
($m$ and $n$ are coprimes, independent of
 $L$; \; $\Delta S^{z}_{T}$ is an integer with $|\Delta S^{z}_{T}| \ll L$)
  are continuous as a function of the wavenumber $q$ 
   in the infinite limit($L \rightarrow \infty$),
   except $\Delta S^{z}_{T}=0$.

\begin{proof}
We shall prove for the case $\Delta S^{z}_T = \pm 1$. Generalization is trivial.

\begin{enumerate}
 \item{$n=1$ case}

      We set the left hand side of equation
      (\ref{eq:Energy-Spectrum-Inequality}) of theorem 2 as
\begin{equation}
\delta^{2} E(q)
\equiv
 E(S^{z}_{T};  q-2\pi /L)
+E(S^{z}_{T};  q+2\pi /L)
- 2 E(S^{z}_{T};  q).
\label{eq:Energy-Spectrum-Difference}
\end{equation}
If the energy spectrum 
were a step function ($E(S^{z}_{T};  q) = \Theta(q-q_{s})$),
then from the following relations:
\begin{equation}
 \delta^{2} E(q) =
\begin{cases}
0 &  |q - q_{s}| \ge 4\pi/L, \\
1/2 &  q = q_{s} -2\pi/L, \\
0 &  q = q_{s}, \\
-1/2 &  q = q_{s} +2\pi/L
\end{cases}
\end{equation}
the inequality (\ref{eq:Energy-Spectrum-Inequality}) could not be satisfied.

This result denies the step discontinuity.
Consequently, possibilities for the essential discontinuity and the infinite
discontinuity are denied.

Therefore, the lowest energy spectrum is continuous.

\item{$n=2$ case}

     Using theorem 2 twice, we obtain
\begin{equation}
 E(S^{z}_{T};  q-4\pi /L)
+E(S^{z}_{T};  q+4\pi /L)
- 2 E(S^{z}_{T};  q)
\le O(1/L),
\end{equation}
therefore, we can prove the continuity similarly to the $n=1$ case.

 \item{$n$ general}
      
      One can prove the continuity similarly.
\end{enumerate}
\end{proof}
  \end{theorem}

[Remark]
One cannot prove the continuity of the lowest energy spectrum in
the $S^{z}_{T}=(S-m/n)L$ subspace.

\medskip

[Corollary]
Although the lowest energy spectra of
$\Delta S^{z}_T =\pm 1,\pm 2, \cdots$
are continuous,
the derivative of the spectra may be discontinuous.
For example, a cusp-like behavior is possible
\begin{equation}
 E(S^{z}_{T};  q)=
  \begin{cases}
v_{1} (q-q_{c}) &  \text{for }  q \ge q_{c}, \\
v_{2} (q-q_{c}) &  \text{for }   q < q_{c}, 
\end{cases}
\end{equation}
in the neighborhood of $q_{c}$.
In this case, from the inequality 
(\ref{eq:Energy-Spectrum-Inequality2}),
the following restriction holds:
\begin{equation}
 (v_{1} - v_{2}) \le \frac{O(1)}{2\pi |S^{z}_{T}|}.
\end{equation}

\subsection{Periodicity of energy spectra}

For special values of the magnetization, 
the wave number change 
of (\ref{eq:Wave-number-eigenstate-of-twisting})
may return to the original wave number in finite times.
Then using the inequality 
(\ref{eq:Energy-Spectrum-Inequality})
several times,
we can show the periodicity of the lowest energy spectra.
This includes the $S^{z}_{T}=0$ case of the original LSM
\cite{Lieb-Schultz-Mattis-1961}
or \cite{Affleck-Lieb-1986}.

\begin{theorem}
(Extension of Oshikawa-Yamanaka-Affleck (OYA) theorem
\cite{Oshikawa-Yamanaka-Affleck-1997}
)

The lowest energy spectra in the subspace
of $S^{z}_{T}=(S-m/n)L$ ($m$ and $n$ are coprimes)
 are periodic with $q\rightarrow q+2\pi/n$
 in the infinite limit:
 \begin{equation}
  |E(S^{z}_{T};  q) - E(S^{z}_{T};  q+2\pi/n) | \le O(1/L).
 \end{equation}
 
\begin{proof}
 
\begin{enumerate}
 \item For the $n=2$ case (or $S-S^{z}_{T}/L=1/2,3/2,\cdots$)

From theorem 2, we obtain
\begin{align}
& E(S^{z}_{T};  q-\pi)
+E(S^{z}_{T};  q +\pi)
- 2 E(S^{z}_{T};  q)
\notag \\
&=
2( E(S^{z}_{T};  q +\pi)
-  E(S^{z}_{T};  q))
\le O(1/L),
\end{align}
where we used
$E(S^{z}_{T};  q' +2\pi)=E(S^{z}_{T}; q') $.
       
Conversely, we can show
\begin{align}
 E(S^{z}_{T};  q)
- E(S^{z}_{T};  q +\pi)
\le O(1/L),
\end{align}
therefore
\begin{equation}
  |E(S^{z}_{T};  q) - E(S^{z}_{T};  q+\pi) | \le O(1/L).
 \end{equation}

 \item For the  $n=3$ case (or $S-S^{z}_{T}/L=1/3,2/3,\cdots$)

       From theorem 2, we obtain
\begin{align}
 E(S^{z}_{T};  q+2\pi/3)
+E(S^{z}_{T};  q-2\pi/3)
- 2 E(S^{z}_{T};  q)
 \le O(1/L).
 \label{eq:OYA-period3-1}
\end{align}

Secondly, applying theorem 2 to the lowest energy state with $q+2\pi /3$, we obtain
\begin{align}
& E(S^{z}_{T};  q+4\pi/3)
+E(S^{z}_{T};  q)
- 2 E(S^{z}_{T};  q+2\pi/3 )
\notag \\
 &= E(S^{z}_{T};  q-2\pi/3)
+E(S^{z}_{T};  q)
- 2 E(S^{z}_{T};  q+2\pi/3 )
\le O(1/L),
 \label{eq:OYA-period3-2}
\end{align}
where we used
$E(S^{z}_{T};  q' +2\pi)=E(S^{z}_{T};  q' ) $.

Thirdly, applying theorem 2 to the lowest energy state with $q-2\pi /3$, we obtain
\begin{align}
& E(S^{z}_{T};  q)
+E(S^{z}_{T};  q-4\pi/3)
- 2 E(S^{z}_{T};  q-2\pi/3 )
\notag \\
 &= E(S^{z}_{T};  q)
+E(S^{z}_{T};  q+2\pi/3)
- 2 E(S^{z}_{T};  q-2\pi/3 )
\le O(1/L).
 \label{eq:OYA-period3-3}
\end{align}

Combining
       (\ref{eq:OYA-period3-1}) $\times 2$ +   (\ref{eq:OYA-period3-3}),
       we obtain
\begin{equation}
 E(S^{z}_{T};  q+2\pi/3) - E(S^{z}_{T};  q) \le O(1/L)
\end{equation}       
On the other hand, from
       (\ref{eq:OYA-period3-2}) $\times 2$ +   (\ref{eq:OYA-period3-3}),
       we obtain
       \begin{equation}
 E(S^{z}_{T};  q) - E(S^{z}_{T};  q+2\pi/3) \le O(1/L)
       \end{equation}       

Therefore,
\begin{equation}
| E(S^{z}_{T};  q)
-  E(S^{z}_{T};  q+2\pi/3)|
\le O(1/L).
\end{equation}

\item
     For general case ($m,n$), we can prove similarly. 

\end{enumerate}
\end{proof}
\end{theorem}

\medskip

\begin{theorem}
 (Second extension of OYA theorem)

In the infinite limit, 
the lowest energy spectra in the subspace 
 $ S^{z}_{T} = L(S-m/n) +\Delta S^{z}_{T}$
 ( $\Delta S^{z}_{T}$ is an integer with  $|\Delta S^{z}_{T}| \ll L$ )
are periodic with $q \rightarrow q+2\pi/n$.
 \begin{equation}
  |E(S^{z}_{T};  q) - E(S^{z}_{T};  q+2\pi/n) | \le O(1/L).
 \end{equation}
 
\begin{proof}

 \begin{enumerate}
 \item 
For the $\Delta S^{z}_{T}=1$ and $(m,n)=(1,2)$  case

       From theorem 2 we obtain
\begin{align}
 &E(S^{z}_{T}; q-\pi +2\pi/L)
  + E(S^{z}_{T}; q+\pi -2\pi/L)
  - 2E(S^{z}_{T}; q) 
\notag \\
 &=
 E(S^{z}_{T}; q+\pi +2\pi/L)
  + E(S^{z}_{T}; q+\pi -2\pi/L)
  - 2E(S^{z}_{T}; q) \le O(1/L)
\end{align}
where we have used
$E(S^{z}_{T};  q' +2\pi)=E(S^{z}_{T};  q' ) $.
In addition, from theorem 3, the lowest energy spectrum 
is a continuous function of $q$
\begin{equation}
| E(S^{z}_{T}; q' \pm 2\pi/L)
  - E(S^{z}_{T}; q')| \le O(1/L)
\end{equation}
Therefore, we obtain
\begin{equation}
	 E(S^{z}_{T}; q +\pi)
  - E(S^{z}_{T}; q) \le O(1/L).
\end{equation}
Conversely, we can show
\begin{equation}
  E(S^{z}_{T}; q) - E(S^{z}_{T}; q +\pi)
  \le O(1/L)
\end{equation}
       In summary
      \begin{equation}
  |E(S^{z}_{T}; q) - E(S^{z}_{T}; q +\pi)|
  \le O(1/L).
\end{equation} 

 \item
For general $\Delta S^{z}_{T}$ and $(m,n)$ cases, 
one can prove
\begin{equation}
  |E(S^{z}_{T};  q) - E(S^{z}_{T};  q+2\pi/n) | \le O(1/L),
\end{equation}
similarly as above and theorem 4.

\end{enumerate}

\end{proof}
\end{theorem}

[Remark]
\begin{figure}[h]
 \includegraphics[width = 40mm]{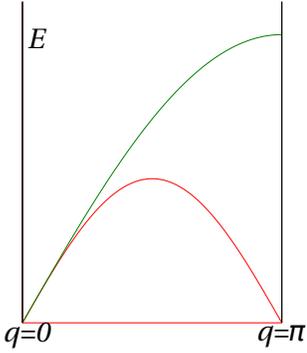}
 \caption{Elementary excitation of S=1/2 Heisenberg chain}
\label{fig:Elementary-excitation-S=1/2-chain}
\end{figure}

The periodicity of theorem 4 and 5 is applied only for the lowest energy spectrum in
      $S^{z}_{T}$, not for higher excitations.
      For example, $S_T=1$ excitation for the S=1/2 Heisenberg model,
      the lower bound behaves as \cite{des-Cloizeaux-Pearson-1962}
      \begin{equation}
       E=\frac{J \pi}{2}\sin{q},
      \end{equation}
      above them there is a continuum of states
      bounded above by \cite{Yamada-1969}
      \begin{equation}
       E=J \pi\sin{\frac{q}{2}}.
      \end{equation}
      (See Fig. \ref{fig:Elementary-excitation-S=1/2-chain})

 \subsection{Correlation}

In the case where $S$ is an integer and $S^{z}_{T}=0$,
one cannot decide whether the ground state is gapless or not,
with the LSM theorem \cite{Affleck-Lieb-1986}.
 In general $S-S^{z}_{T}/L$ integer, there is a similar statement 
 \cite{Oshikawa-Yamanaka-Affleck-1997}. 
  
  However in such a situation, there is a restriction on the expectation
value of the matrix elements (or correlations).

  \medskip
 \begin{theorem} 
When $S^{z}_{T}=(S-m)L$ ($m$: integer), the following inequality holds.
\begin{align}
&
\langle  S^{z}_{T};  q | 
((\hat{U}^{\rm tw}_{2\pi})^{\dagger} \hat{H} \hat{U}^{\rm tw}_{2\pi}  
+(\hat{U}^{\rm tw}_{-2\pi})^{\dagger} \hat{H} \hat{U}^{\rm tw}_{-2\pi} 
- 2 \hat{H}
 )
| S^{z}_{T};  q \rangle
 \ge 0.
\label{eq:Expectation-Inequality}
\end{align}

\proof

 We can show  (\ref{eq:Expectation-Inequality}), 
 using the inequality (\ref{eq:Energy-Spectrum-Inequality2}) 
 in theorem 2. 
 \end{theorem}

For example (\ref{eq:Hamiltonian-generalized-XXZ}) with
$J(r)=\delta_{r,1}$,
this means
\begin{equation}
       \langle S^{z}_{T}=(S-m)L;  q | 
	\hat{S}^{x}_{j}\hat{S}^{x}_{j+1} | S^{z}_{T}=(S-m)L;  q \rangle \le 0.
\end{equation}

\section{Discrete symmetries}

In this section,
in addition to the translational symmetry,
we will discuss discrete symmetries
(space inversion, spin reversal).

\subsection{Symmetry in the lowest energy spectrum} 

According to the Bethe ansatz for the S=1/2 XXZ spin chain,
the lowest energy spectrum is symmetric with $q=\pi/2$.
This can be proven even when there is no exact result.

When the Hamiltonian is invariant under the space inversion (site
parity or link parity), besides the translational invariance,
there is $E(S^{z}_{T}; -q)=E(S^{z}_{T}; q)$
symmetry in the energy spectra.

\medskip

 \begin{theorem}(Third extension of OYA theorem)

  In the infinite limit, 
the lowest energy spectra of
$S^{z}_{T} = (S-m/n)L+\Delta S^{z}_{T}; (\Delta S^{z}_{T}=0,\pm 1,\pm2,\cdots)$
are symmetric with respect to $q=\pi/n$,

\begin{equation}
 |E(S^{z}_{T};q)-E(S^{z}_{T};2\pi/n -q)| \le O(1/L).
\end{equation}  
  \begin{proof}

Combining the theorem 5 and $E(S^{z}_{T}; -q')=E(S^{z}_{T}; q')$,
 we can prove this theorem.
  \end{proof}

 \end{theorem}

\medskip

[Remark]
\begin{itemize}
 \item 
From the above theorem, 
for the $S$ half-integer case, the lowest energy spectrum of 
$S^{z}_{T} =0,\pm 1,\pm 2,\cdots $  is symmetric with respect to
$q=\pi/2$ in the infinite limit
\cite{Kolb-1985,Fath-Solyom-1993}
 \item
This symmetry does not hold for higher energy spectra than the lowest
spectrum.
      (See Fig. \ref{fig:Elementary-excitation-S=1/2-chain})

\end{itemize}

 \subsection{Discrete Symmetries and twisting operator}

 \subsubsection{Site parity}
\begin{lemma}
(Space inversion and twisting operators)

\begin{equation}
 \hat{P}\hat{U}^{\rm tw}_{\pm 2\pi} = \hat{U}^{\rm tw}_{\mp 2\pi} \hat{P}
\label{eq:TBC-Parity-2pi}.
\end{equation}

\begin{proof}
 
We can obtain
\begin{align}
\hat{P}\hat{U}^{\rm tw}_{\pm 2 \pi}\hat{P}
&= \exp\left(\mp \frac{2\pi i}{L}\sum_{j=1}^{L}j (\hat{S}^{z}_{L-j}-S) \right)
\notag \\
&= \exp\left(\mp\frac{2\pi i}{L}\sum_{j=0}^{L-1}(L-j) (\hat{S}^{z}_{j}-S) \right)
\notag \\
&= 
\hat{U}^{\rm tw}_{ \mp 2\pi}
\exp(\mp 2\pi i (\hat{S}^{z}_{T}-SL))
\exp(\mp 2\pi i (\hat{S}^{z}_{L}-S)),
\label{eq:TBC-Parity}
\end{align}
where we used $\hat{S}^z_{0} = \hat{S}^z_{L} $.
Combining this relation with
the fact 
that the eigenvalues of
$\hat{S}^{z}_{T}-SL$ and
$\hat{S}^{z}_{L}-S$ 
are integers,
we can obtain (\ref{eq:TBC-Parity-2pi}).
\end{proof}
\end{lemma}

\subsubsection{Spin reversal symmetry}

\begin{lemma}
($\pi$ rotation around the $y$-axis and twisting operator)

\begin{equation}
 \hat{U}^{y}_{\pi} \hat{U}^{\rm tw}_{\pm 2\pi} 
= (-1)^{2S} \hat{U}^{\rm tw}_{\mp 2\pi} \hat{U}^{y}_{\pi}.
\label{eq:TBC-y-pi-inversion}
\end{equation}

\begin{proof}
 
\begin{align}
(\hat{U}^{y}_{\pi})^{\dagger}\hat{U}^{\rm tw}_{\pm 2 \pi} \hat{U}^{y}_{\pi}
&= \exp\left(\mp \frac{2\pi i}{L}\sum_{j=1}^{L}j (-\hat{S}^{z}_{j}-S) \right)
\notag \\
&= \exp\left(\pm \frac{2\pi i}{L}\sum_{j=0}^{L} j (\hat{S}^{z}_{j}-S)
 \right)
\exp\left( \pm \frac{4\pi i}{L} S\sum_{j=1}^{L} j \right)
\notag \\
&= 
\hat{U}^{\rm tw}_{ \mp 2\pi}
\exp ( \pm 2\pi i S(L+1) )
\end{align}
Combining this relation with
the fact $SL$ is an integer, we can show equation (\ref{eq:TBC-y-pi-inversion}). 
\end{proof}
\end{lemma}

\subsubsection{Link Parity}

Since the twisting operator defined (\ref{eq:Twisting-Operator})
and the link parity $\hat{P}_{\rm link}$
are not compatible with lemma 2,
we should define another type twisting unitary operator as
equation (17) and appendix C in \cite{Nomura-Kitazawa-1998}
\begin{equation}
 \hat{U}^{twl}_{\pm 2\pi}
  \equiv
  \exp \left(\mp \frac{2\pi i}{L} \sum_{j=1}^{L} \left( j-\frac{1}{2} \right) (\hat{S}^{z}_{j}
  -S)\right).
\end{equation}

\begin{lemma}
 One can show the relation
\begin{equation}
 \hat{P}_{\rm link}  \hat{U}^{twl}_{\pm 2\pi} \hat{P}_{\rm link} 
  = \hat{U}^{twl}_{\mp 2\pi}.
\end{equation}
  \begin{proof}
   \begin{align}
    \hat{P}_{\rm link}\hat{U}^{twl}_{\pm 2 \pi}\hat{P}_{\rm link}
&= \exp\left(\mp \frac{2\pi i}{L}\sum_{j=1}^{L}(j-1/2) (\hat{S}^{z}_{L-j+1}-S) \right)
\notag \\
&= \exp\left(\mp\frac{2\pi i}{L}\sum_{j=1}^{L}(L-j+1-1/2) (\hat{S}^{z}_{j}-S) \right)
\notag \\
&= 
\hat{U}^{twl}_{ \mp 2\pi}
    \exp(\mp 2\pi i (\hat{S}^{z}_{T}-SL))
    \notag \\
&=\hat{U}^{twl}_{ \mp 2\pi}.
   \end{align}

  \end{proof}

\end{lemma}
However, the corresponding relation for lemma 3 should be changed as
\begin{lemma}
\begin{equation}
 \hat{U}^{y}_{\pi} \hat{U}^{twl}_{\pm 2\pi} 
= \hat{U}^{twl}_{\mp 2\pi} \hat{U}^{y}_{\pi}.
\end{equation}
 \begin{proof}
 
\begin{align}
(\hat{U}^{y}_{\pi})^{\dagger}\hat{U}^{twl}_{\pm 2 \pi} \hat{U}^{y}_{\pi}
&= \exp\left(\mp \frac{2\pi i}{L}\sum_{j=1}^{L}(j-1/2) (-\hat{S}^{z}_{j}-S) \right)
\notag \\
&= \exp\left(\pm \frac{2\pi i}{L}\sum_{j=0}^{L} (j-1/2) (\hat{S}^{z}_{j}-S)
 \right)
\exp\left( \pm \frac{4\pi i}{L} S\sum_{j=1}^{L} (j-1/2) \right)
\notag \\
&= 
\hat{U}^{twl}_{ \mp 2\pi}
\exp ( \pm 2\pi i SL )
\notag \\
&= 
\hat{U}^{twl}_{ \mp 2\pi}
\end{align}
\end{proof}
  
\end{lemma}

\subsection{Combination of discrete symmetries}

\begin{theorem}(Corresponding to the equation (29) of  \cite{Affleck-Lieb-1986})

On the combination $\hat{P}\hat{U}^{y}_{\pi} $,
the next relation holds
\begin{equation}
 (\hat{P}\hat{U}^{y}_{\pi} )\hat{U}^{\rm tw}_{\pm 2\pi}
= (-1)^{2S} \hat{U}^{\rm tw}_{\pm 2\pi} (\hat{P}\hat{U}^{y}_{\pi} )
\end{equation}
\begin{proof}
Combining lemmas 2 and 3, one can prove this.
\end{proof}

\end{theorem}

\begin{theorem}

On the combination $\hat{P}_{\rm link}\hat{U}^{y}_{\pi} $,
the next relation holds
\begin{equation}
 (\hat{P}_{\rm link}\hat{U}^{y}_{\pi} )\hat{U}^{twl}_{\pm 2\pi}
= \hat{U}^{twl}_{\pm 2\pi} (\hat{P}_{\rm link}\hat{U}^{y}_{\pi} )
\end{equation}
\begin{proof}
Combining lemmas 4 and 5, one can prove this.
\end{proof}

\end{theorem}

For example, 
in the spin half-integer case,
since the eigenstate
$|\psi_{0} \rangle \equiv | S^{z}_{T}=0;  q=0 \rangle $
satisfies
$\hat{P}  \hat{U}^{y}_{\pi}|\psi_{0}\rangle = \pm|\psi_{0} \rangle$,
therefore from theorem 9 we obtain 
$(\hat{P} \hat{U}^{y}_{\pi}) \hat{U}^{\rm tw}_{2 \pi}|\psi_{0} \rangle =
\mp\hat{U}^{\rm tw}_{2 \pi}|\psi_{0} \rangle$,
or
$\langle \psi_{0} |  \hat{U}^{\rm tw}_{2 \pi}|\psi_{0} \rangle=0$.

\section{Consideration of several models}

In this section we consider several models
other than (\ref{eq:Hamiltonian-generalized-XXZ}).
Since the lattice structure or the symmetries of them
may be different from the model (\ref{eq:Hamiltonian-generalized-XXZ}),
we should slightly modify the statements of theorems 1-9 in several cases.

\subsection{XXZ spin chain with next-nearest-neighbor interaction}
We consider an XXZ spin chain with next-nearest-neighbor (NNN) interaction:
\begin{align}
 \hat{H} &= \sum_{j=1}^{L} 
\left(
\frac{1}{2}(\hat{S}^{+}_{j}
  \hat{S}^{-}_{j+1}+\hat{S}^{-}_{j} \hat{S}^{+}_{j+1}  )
  +\Delta  \hat{S}^{z}_{j} \hat{S}^{z}_{j+1}
\right) 
+\alpha 
\sum_{j=1}^{L} 
\left(
 \frac{1}{2}(\hat{S}^{+}_{j}
  \hat{S}^{-}_{j+2}+\hat{S}^{-}_{j} \hat{S}^{+}_{j+2}  )
  +\Delta  \hat{S}^{z}_{j} \hat{S}^{z}_{j+2}
\right).
\label{eq:Hamiltonian-NNN-XXZ}
\end{align}
This Hamiltonian is invariant with $z$-axis rotation,
 translation, space inversion(link parity, site parity),
 and $\pi$ rotation around the $y$-axis.

Theorems 1-9 hold.
When $\alpha >0$ with frustration, the MLM theorem
does not hold,
therefore the uniqueness of the ground state may be broken.

In the S=1/2 case, there are several exact results.
At $\alpha=0$ there is the exact results by Bethe ansatz
\cite{Bethe-1931,des-Cloizeaux-Pearson-1962,Yang-Yang-1966,Yamada-1969},
where energy spectra are known,
and these results are consistent with our theorems.
At $\alpha=1/2$ there is another type exact result
\cite{Majumdar-Ghosh-1969},
where two-fold exactly degenerate ground states $q=0,\pi$ exist even for
finite size.
Unfortunately, there is no exact result for excitation spectra at
$\alpha=1/2$.

\subsection{Bilinear-biquadratic (BLBQ) spin chain}

We consider a bilinear-biquadratic (BLBQ) spin chain :
\begin{align}
 \hat{H} &= \sum_{j=1}^{L} 
\left(
 \hat{\bm{S}}_{j}\cdot \hat{\bm{S}}_{j+1}
+
\alpha
(\hat{\bm{S}}_{j}\cdot \hat{\bm{S}}_{j+1})^2
\right).
\label{eq:Hamiltonian-BLBQ}
\end{align}
This Hamiltonian is invariant with $z$-axis rotation,
 translation, space inversion(link parity, site parity),
 and $\pi$ rotation around the $y$-axis.

 Note that in this case although there appear terms such that
\begin{equation}
 \hat{S}_{j}^{+}\hat{S}_{j+1}^{-}\hat{S}_{j}^{+}\hat{S}_{j+1}^{-}, \;
  \hat{S}_{j}^{+}\hat{S}_{j+1}^{-}\hat{S}_{j}^{-}\hat{S}_{j+1}^{+}, \;
  \hat{S}_{j}^{z}\hat{S}_{j+1}^{z}\hat{S}_{j}^{+}\hat{S}_{j+1}^{-},\cdots,
  \end{equation}
 it is straightforward to show the relation
 (\ref{eq:Energy-Spectrum-Inequality2}).
Therefore, theorems 1-9 hold.
When $\alpha >0$ with frustration, the MLM theorem
does not hold,

In the case of S=1,
there are exact solutions by Bethe ansatz at
$\alpha=1$ 
\cite{Uimin-Lai-Sutherland}
where gapless excitations are at $q=0,\pm 2\pi/3$, 
and
$\alpha=-1$ 
\cite{Takhtajan-Babujian}
where gapless excitations are at $q=0, \pi$.
At 
$\alpha=1/3$ there is another type exact result
\cite{Affleck-Kennedy-Lieb-Tasaki-1988}.
A unique ground state with an energy gap has been proved.
These results are consistent with theorems 1-9.

\subsection{XXZ spin chain with staggered field}
\begin{align}
 \hat{H} &= \sum_{j=1}^{L} 
\left(
\frac{1}{2}(\hat{S}^{+}_{j}
  \hat{S}^{-}_{j+1}+\hat{S}^{-}_{j} \hat{S}^{+}_{j+1}  )
  +\Delta  \hat{S}^{z}_{j} \hat{S}^{z}_{j+1}
\right) 
+\delta \sum_{j=1}^{L} (-1)^{j}  \hat{S}^{z}_{j}
\end{align}
This Hamiltonian is invariant with $z$-axis rotation,
translation by two sites, space inversion(site parity).

Theorems 1-7 hold with several changes.

Considering two spins in one unit cell,
we should use the translation operator by two sites
and rewrite the lemma 1 as:
\begin{align}
 (\hat{U}_{\rm trl})^{2} \hat{U}^{\rm tw}_{2\pi}
=  \hat{U}^{\rm tw}_{2\pi} (\hat{U}_{\rm trl})^{2}
 \exp\left(-\frac{4 \pi i}{L}\hat{S}^{z}_{T}
 \right),
\end{align}
where we use $2S$ as the integer.

\subsection{XXZ spin chain with bond-alternation}

\begin{align}
 \hat{H} &= \sum_{j=1}^{L} 
(1+\delta (-1)^{j})
\left(
\frac{1}{2}(\hat{S}^{+}_{j}
  \hat{S}^{-}_{j+1}+\hat{S}^{-}_{j} \hat{S}^{+}_{j+1}  )
  +\Delta  \hat{S}^{z}_{j} \hat{S}^{z}_{j+1}
\right) 
\end{align}
This Hamiltonian is invariant with $z$-axis rotation,
translation by two sites, space inversion(link parity),
 and $\pi$ rotation around the $y$-axis.

Theorems 1-7 and 9 hold with several changes.
Lemma 1 becomes
\begin{align}
 (\hat{U}_{\rm trl})^{2} \hat{U}^{twl}_{2\pi}
=  \hat{U}^{twl}_{2\pi} (\hat{U}_{\rm trl})^{2}
 \exp\left(-\frac{4 \pi i}{L}\hat{S}^{z}_{T}
 \right).
\end{align}

In the case of $\delta=0, \Delta \gg 0 $ the ground state is N\'eel
state, that is, two-fold degenerate ($q=0,\pi$).
Therefore, when $|\delta| \ll 1$,
corresponding to the folding of the Brillouin zone,
the lowest ground state is two-fold degenerate in the subspace
$S^{z}_{T}=0;  q=0$ but different discrete symmetries
($\hat{P}_{\rm link}=1; \hat{U}^{y}_{\pi}=1$ versus
$\hat{P}_{\rm link}=-1; \hat{U}^{y}_{\pi}=-1$ ),
which is robust against perturbation $\delta \neq 0$.

\subsection{Dzyaloshinskii-Moriya type interaction}

The Dzyaloshinskii-Moriya type interaction
\cite{Dzyaloshinskii-1958,Moriya-1960}
\begin{equation}
 \hat{H}_{DM}
  = \sum_{j}  (\hat{\bm{S}}_{j} \times \hat{\bm{S}}_{j+1})^{z}
  = \frac{i}{2}\sum_{j}  (\hat{S}_{j}^{+} \hat{S}_{j+1}^{-} -
  \hat{S}_{j}^{-} \hat{S}_{j+1}^{+})
  \label{eq:Dzyaloshinskii-Moriya-interaction}
\end{equation}
is U(1) and translational invariant, but antisymmetric for the
space inversion.

One can easily show  (\ref{eq:Energy-Spectrum-Inequality2}),
therefore theorems 1-6 hold. 
Note that in this case there is not the $q \rightarrow -q $ symmetry
in the dispersion curve,
which may be related with the spin spiral ordering.

\subsection{Nonsymmetric spin ladder}

\begin{figure}[h]
 \includegraphics[width = 40mm]{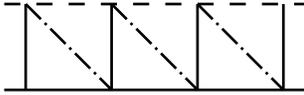}
 \caption{Spin Ladder with nonsymmetric interaction}
\label{fig:Ladder-nonsymmetirc}
\end{figure}

We can consider a spin ladder model with nonsymmetric interactions shown in
Fig.\ref{fig:Ladder-nonsymmetirc}
with U(1) symmetry.

One can easily show  (\ref{eq:Energy-Spectrum-Inequality2}),
therefore theorems 1-6 hold.

\section{Spectra minima from the MLM theorem}

For finite magnetization  
 $S^{z}_{T} = (S-m/n)L,(S-m/n)L \pm 1,\cdots$ cases,
 from theorems 4,5 and 7 there remain two possibilities
 for the minimum of energy spectrum; $q=0$ or $q=\pi$.
Using only the symmetries such as U(1), translation, space inversion or
 spin reversal,
 one cannot conclude which is appropriate. 

Nevertheless, in the nonfrustrating case,
using the MLM theorem
\cite{Marshall-1955,Lieb-Mattis-1962},
one can distinguish the above two situations.
Using the alternating operator:
\begin{equation}
 \hat{U}^{\rm alt} \equiv \exp \left( \pi i \sum_{j=1}^{L} j (\hat{S}^{z}_{j}
			    -S)\right),
 \label{eq:Alternating-Unitary}
\end{equation}
then we obtain
\begin{align}
(\hat{U}^{\rm alt})^{\dagger} \hat{S}^{\pm}_j \hat{U}^{\rm alt}
=(-1)^{j} \hat{S}^{\pm}_j, 
\quad
 (\hat{U}^{\rm alt})^{\dagger} \hat{S}^{z}_j \hat{U}^{\rm alt}
=  \hat{S}^{z}_j. 
\end{align}
For example, considering the NNN spin chain
 (\ref{eq:Hamiltonian-NNN-XXZ}) with $\alpha \le 0$,
the unitary transformation 
with this operator is
\begin{align}
 (\hat{U}^{\rm alt})^{\dagger} \hat{H} \hat{U}^{\rm alt}
 &= \sum_{j=1}^{L} 
\left( - \frac{1}{2}(\hat{S}^{+}_{j}
  \hat{S}^{-}_{j+1}+\hat{S}^{-}_{j} \hat{S}^{+}_{j+1}  )
  +\Delta  \hat{S}^{z}_{j} \hat{S}^{z}_{j+1}
 \right)\notag \\
&+\alpha \sum_{j=1}^{L} 
\left( \frac{1}{2}(\hat{S}^{+}_{j}
  \hat{S}^{-}_{j+2}+\hat{S}^{-}_{j} \hat{S}^{+}_{j+2}  )
  +\Delta  \hat{S}^{z}_{j} \hat{S}^{z}_{j+2}
 \right),
 \label{eq:XXZ-PF}
\end{align}
therefore the off-diagonal elements of the Hamiltonian become nonpositive,
and one can use the Perron-Frobenius theorem
(diagonal elements can be adjusted by adding
a scalar multiple of the identity operator).

When system size $L$ is finite,
from the Perron-Frobenius theorem,
the lowest energy state of (\ref{eq:XXZ-PF}) in each $S^{z}_{T}$ subspace
is unique with the wavenumber $q=0$. 
Returning to the model 
 (\ref{eq:Hamiltonian-NNN-XXZ}) with $\alpha \le 0$
by the unitary operator (\ref{eq:Alternating-Unitary}),
the minimum of the energy spectrum is located 
at $q=0$ in the case $SL - S^{z}_{T}$ even integer,
whereas the minimum of spectrum is at $q=\pi$ in the case $SL - S^{z}_{T}$ odd integer. 
Especially when $L=2n$, energy spectrum in $S^{z}_{T} = (S-m/n)L$ subspace
has a minimum at $q=0$.
Similar discussion can be applied for the BLBQ model (\ref{eq:Hamiltonian-BLBQ}) with
$\alpha \le 0$.

Although for frustrating case, the above consideration may become
ineffective,
one may expect such an even-odd difference in the $SL - S^{z}_{T}$ subspace.

\section{Comparison with previous works}

In this section we compare previous works and our theorems.

\subsection{Problem of the original LSM discussion}

For the S=1/2 XXZ model
\begin{align}
 \hat{H} &= \sum_{j=1}^{L} 
\left( \frac{1}{2}(\hat{S}^{+}_{j}
  \hat{S}^{-}_{j+1}+\hat{S}^{-}_{j} \hat{S}^{+}_{j+1}  )
  +\Delta  \hat{S}^{z}_{j} \hat{S}^{z}_{j+1}
\right),
\label{eq:Hamiltonian-XXZ}
\end{align}
Lieb, Schultz and Mattis \cite{Lieb-Schultz-Mattis-1961}
discussed as follows:
\begin{align}
& (\hat{U}^{\rm tw}_{2\pi})^{\dagger} \hat{H} \hat{U}^{\rm tw}_{2\pi} -\hat{H}
\notag \\
&=\sum_{j=1}^{L} 
\frac{1}{2}
\left(
\hat{S}^{+}_{j} \hat{S}^{-}_{j+1}(\exp(-i 2\pi/L)-1)
+\hat{S}^{-}_{j} \hat{S}^{+}_{j+1}(\exp(+i 2\pi/L)-1) 
\right)
\notag \\
&=
\frac{i}{2}\sin\left( \frac{2 \pi}{L} \right)
\sum_{j=1}^{L} 
(-\hat{S}^{+}_{j} \hat{S}^{-}_{j+1}
+\hat{S}^{-}_{j} \hat{S}^{+}_{j+1})
+ 
\frac{1}{2}\left(\cos \left( \frac{2 \pi}{L} \right)-1 \right)
\sum_{j=1}^{L} 
(\hat{S}^{+}_{j} \hat{S}^{-}_{j+1}
+\hat{S}^{-}_{j} \hat{S}^{+}_{j+1}
), 
\end{align}
and
\begin{align}
\langle \Psi_{0} |\sum_{j=1}^{L} 
(-\hat{S}^{+}_{j} \hat{S}^{-}_{j+1}
 +\hat{S}^{-}_{j} \hat{S}^{+}_{j+1})| \Psi_{0} \rangle
= 
\langle \Psi_{0} | \left[ \left(\sum_{j=1}^{L} 
j \hat{S}^{z}_{j}\right), \hat{H} \right] | \Psi_{0} \rangle=0,
\end{align}
where $|\Psi_{0}\rangle$ is the ground state.
The remaining reasoning is similar to ours.

However, we think there is a technical problem in their method.
For later convenience, we rewrite
 \begin{align}
  &(\hat{U}^{\rm tw}_{2\pi})^{\dagger} \hat{H} \hat{U}^{\rm tw}_{2\pi} -\hat{H}
=
  \frac{1}{2}( \widehat{ST} +\widehat{AST} ),
  \notag \\
&   \widehat{ST}
\equiv
(\hat{U}^{\rm tw}_{2\pi})^{\dagger} \hat{H} \hat{U}^{\rm tw}_{2\pi}
  + (\hat{U}^{\rm tw}_{-2\pi})^{\dagger} \hat{H} \hat{U}^{\rm tw}_{-2\pi}
-2\hat{H},
\notag \\
&  \widehat{AST}
\equiv
(\hat{U}^{\rm tw}_{2\pi})^{\dagger} \hat{H} \hat{U}^{\rm tw}_{2\pi}
  - (\hat{U}^{\rm tw}_{-2\pi})^{\dagger} \hat{H} \hat{U}^{\rm tw}_{-2\pi}.
  \label{eq:Even-Odd-Decomposition-LSM}
 \end{align}
 
For (\ref{eq:Hamiltonian-XXZ}) it is easy to show
 that the second term of
 (\ref{eq:Even-Odd-Decomposition-LSM})
 becomes a commutator form of the Hamiltonian
 \begin{equation}
   \widehat{AST}
  = i \sin \left(\frac{2\pi}{L} \right) \left[ \left(\sum_{j=1}^{L} j \hat{S}^{z}_{j}\right), \hat{H}\right].
 \end{equation}
Whereas for the NNN Hamiltonian (\ref{eq:Hamiltonian-NNN-XXZ}),
there appears an additive term to the commutator:
\begin{align}
 \widehat{AST} 
  &  = i \sin \left( \frac{2\pi}{L} \right) \left[ \left(\sum_{j=1}^{L} j \hat{S}^{z}_{j}\right),
  \hat{H}\right]
  \notag \\
  &  + 2\alpha i \sin\left(\frac{2\pi}{L}\right)\left( \cos\left(\frac{2\pi}{L}\right)-1\right)\sum_{j=1}^{L}(-\hat{S}^{+}_{j} \hat{S}^{-}_{j+2} + \hat{S}^{-}_{j} \hat{S}^{+}_{j+2})
. 
\end{align}
For the BLBQ model (\ref{eq:Hamiltonian-BLBQ}),
it is cumbersome to derive
 \cite{Affleck-Lieb-1986}
 \begin{equation}
  \widehat{AST}
  = i \sin \left( \frac{2\pi}{L} \right) \left[ \left(\sum_{j=1}^{L} j \hat{S}^{z}_{j}\right), \hat{H}
	    \right] +O(L^{-1}).
 \end{equation}
These calculations,
which become more difficult for complicated models,
are highly model dependent,
which are different from our model independent proof of theorem 2 etc.

\subsection{Problems of Oshikawa-Yamanaka-Affleck discussion}

Oshikawa {\it et al.}
\cite{Oshikawa-Yamanaka-Affleck-1997}
have assumed the space inversion symmetry or the spin reversal symmetry
of the model.
Although they have not presented explicitly the reason,
we guess it would be a workaround to avoid the difficulty of the
previous subsection.
In fact,
when model is space inversion symmetric, from lemma 2 it suffices that
\begin{equation}
 \hat{P} \widehat{AST}\hat{P} = - \widehat{AST},
\end{equation}
or when model is spin reversal symmetric, from lemma 3 it suffices that
\begin{equation}
(\hat{U}^{y}_{\pi})^{\dagger} \widehat{AST}\hat{U}^{y}_{\pi} = - \widehat{AST},
\end{equation}
therefore, it is apparent that
\begin{equation}
  \langle \Psi_{0}| \widehat{AST} | \Psi_{0}\rangle =0,
\end{equation}
where the state
$ \hat{P} | \Psi_{0}\rangle =  | \Psi_{0}\rangle$ or
$ \hat{U}^{y}_\pi | \Psi_{0}\rangle =  | \Psi_{0}\rangle$.

However, there are problems.
 First,
 for the states $|S^{z}_{T}; q\rangle$, the discussion becomes not so simple;
 one should use the combination
 $|S^{z}_{T}; q\rangle+ |S^{z}_{T}; -q\rangle$,
 which leads the three wave numbers inequality 
 (\ref{eq:Energy-Spectrum-Inequality}),
 not the original LSM type.

  Secondly, with the discrete symmetry assumption,
  several models are excluded such as
  the Dzyaloshinskii-Moriya  interaction,
or nonsymmetric spin ladders, or distorted diamond chain models with
  a staggered field.

\section{Discussions on energy spectra}

In this section we illustrate possible energy spectra
from our theorems.

\subsection{Dispersion curve for frustrating system}

First we treat the frustrate systems with space inversion symmetry.
The lowest energy spectrum of $S^{z}_{T}=\pm 1$ is continuous from theorem 3,
and there are symmetry restrictions from theorems 1,5 and 7.

\begin{itemize}
 \item{S: half-integer spin}

Regarding the dispersion curve in $S^{z}_{T}=\pm 1$,
there is the possibility of the four lowest points,
other than the conventional two lowest points $q= 0,\pi$.
(See figure \ref{fig:Spin-half-dispersion})
\begin{figure}[h]
 \includegraphics[width = 80mm]{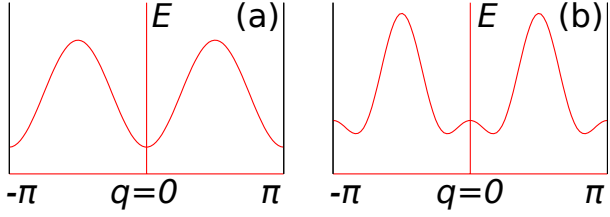}
 \caption{Possible energy spectrum $S=1/2,3/2,\cdots$ 
 (a) two lowest points (b) four lowest points}
 \label{fig:Spin-half-dispersion}
\end{figure}

For example,
when the next-nearest-neighbor interaction $\alpha$ is large enough
in the S=1/2 NNN XXZ spin chain  
(\ref{eq:Hamiltonian-NNN-XXZ}),
dispersion curve may have four minima.

 \item{S: integer spin}

Regarding the dispersion curve in $S^{z}_{T}=\pm 1$,
there is the possibility of the two lowest points($q\neq 0,\pi$) ,
other than the conventional unique lowest point.
(See figure \ref{fig:Spin-one-dispersion})
\begin{figure}[h]
 \includegraphics[width = 80mm]{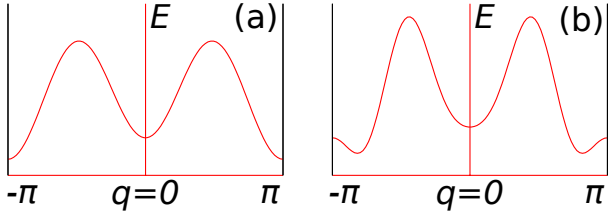}
\caption{Possible energy spectrum for $S=1,2,\cdots$
 (a) unique lowest point (b) two lowest points}
 \label{fig:Spin-one-dispersion}
\end{figure}

For example,
when the biquadratic interaction $\alpha$ is large enough
in the S=1 BLBQ spin chain  
(\ref{eq:Hamiltonian-BLBQ}),
dispersion curve may have two minima.
      
\end{itemize}

\subsection{Magnetization plateaux}

\begin{figure}[h]
 \includegraphics[width = 80mm]{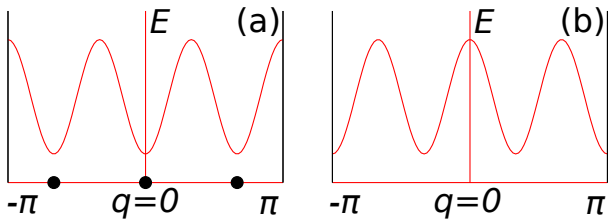}
 \caption{Possible energy spectra 
 (a) $S^{z}_{T}=(S-1/3) L$ (b) $S^{z}_{T}=(S-1/3) L+1$ }
 \label{fig:Magnetization-plateau-dispersion}
 \end{figure}

We consider the energy spectra at magnetization plateaux.
As examples, we treat
the 
$S^{z}_{T}=(S-1/3) L$ case
and
the
$S^{z}_{T}=(S-1/3) L+1$ case.
In these cases, the lowest energy spectra are periodic with
the wavenumber $q \rightarrow q+ 2\pi/3$.
In the $S^{z}_{T}=(S-1/3) L$ subspace,
the lowest energy spectrum may be discontinuous,
and the lowest energy state is located at
$q=0$ 
from the MLM discussion in the section 6 
(figure \ref{fig:Magnetization-plateau-dispersion} (a)).
In the $S^{z}_{T}=(S-1/3) L+1$ subspace,
the lowest energy spectrum must be continuous with $q$
from theorem 3,
and the minimum point is at $q=\pi$
(figure \ref{fig:Magnetization-plateau-dispersion} (b)).

 \subsection{Nonsymmetric dispersion}

\begin{figure}[h]
 \includegraphics[width = 40mm]{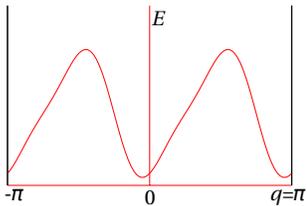}
 \caption{Possible energy spectrum for nonsymmetric model}
 \label{fig:dispersion-nonsymmetirc}
\end{figure}

 We consider energy spectra of
 a model with Dzyaloshinskii-Moriya type interactions
 (\ref{eq:Dzyaloshinskii-Moriya-interaction})
or 
the nonsymmetric spin ladder (see figure \ref{fig:Ladder-nonsymmetirc}).
In this case, energy spectra are periodic from theorem 4 and 5,
whereas they are not symmetric under $q \leftrightarrow -q$
(figure \ref{fig:dispersion-nonsymmetirc}).

 \section{Conclusions}

 We have extended the LSM theorem including the frustrated case,
because we have not used the uniqueness condition of the lowest state in
each $S^{z}_{T}$ subspace.
We have also extended the LSM theorem for nonsymmetric case, 
for example, the Dzyaloshinskii-Moriya interaction.

Regarding the continuity of the lowest energy spectra
in the $S^{z}_{T}=\pm 1,\pm2,\cdots $ subspace,
we have completed insufficient points of previous works
\cite{Fath-Solyom-1993},
since they used inequality relation only once.
Relating to this, Oshikawa \cite{Oshikawa-2000} stated
``that the incommensurate filling gives a gapless spectrum
is empirically recognized more or less'',
and he tried to prove it with topological arguments.
Although his argument is intuitive, it is not convincing.

There are several remarks.
First, on the ordering of the lowest energies of each $S^{z}_{T}$ subspace;
one can say nothing about it from the extended LSM theorem itself.
Secondly, the lowest energy states in the $S^{z}_{T}=(S-m/n)L$ subspace
may be discontinuous as a function of wave number $q$,
which is natural, considering N\'eel or dimer states
or magnetic plateaux.
Thirdly, cautions should be taken on the number of spins in the unit cell.

Finally, the original LSM theorem has also been extended for fermion systems on
the lattice \cite{Yamanaka-Oshikawa-Affleck-1997,Gagliardini-Hass-Rice-1998}.
 It will be interesting to consider our methods for fermion
models with frustration. 

\section*{Acknowledgement}

We would like to thank Hosho Katsura for pointing out
\cite{Hakobyan-2003} and references therein.
K. N. wishes to thank Tohru Koma for the suggestion 
that the space inversion assumption is not necesssary 
for the LSM-type theorem.

\pagebreak


\begin{thebibliography}{99}
\bibitem{Lieb-Schultz-Mattis-1961}
E. Lieb, T. Schultz and D. Mattis:
Annals of Physics,
{\bf 16}, 
(1961), p. 407.
\bibitem{Bethe-1931}
H. Bethe: Z. Phys. {\bf 71} (1931) p.205.
\bibitem{des-Cloizeaux-Pearson-1962}
J. des Cloizeaux and J. J. Pearson:
Phys. Rev. {\bf 128} (1962) p. 2131
\bibitem{Yang-Yang-1966}
C. N. Yang and C. P. Yang: Phys. Rev. {\bf 150} (1966) p. 321.
 \bibitem{Yamada-1969}
	 T. Yamada: Prog. Theor. Phys. {\bf 41} (1969) p.880.
 \bibitem{Majumdar-Ghosh-1969}
C. K. Majumdar and D. K. Ghosh:
 J. Math. Phys. {\bf 10}, (1969), p. 1388 ;
C. K. Majumdar:
J. Phys. C: Solid State Phys. {\bf 3} (1970), p. 911.
\bibitem{Affleck-Kennedy-Lieb-Tasaki-1988}
I. Affleck, T. Kennedy, E. H. Lieb, and H. Tasaki:
    Comm. Math. Phys.
    {\bf 115}, (1988), p.477.
 \bibitem{Marshall-1955}
W. Marshall:
Proc. Roy. Soc., {\bf A232}, (1955), p.48
\bibitem{Lieb-Mattis-1962}
	E. H. Lieb  and D. Mattis: J. Math. Phys.  {\bf 3}, (1962) p.749.
\bibitem{Munro-1976}
	 R. G. Munro: Phys. Rev. B {\bf 13}, (1976) p. 4875.
\bibitem{Affleck-Lieb-1986}
I. Affleck and E. H. Lieb:
Letters in Mathematical Physics. {\bf 12}, (1986) p.57.
\bibitem{Kolb-1985}
M. Kolb:
Phys. Rev. B {\bf 31}, (1985) p. 7494.
\bibitem{Fath-Solyom-1993}
G. F\'ath and J. S\'olyom:
Phys. Rev. B {\bf 47}, (1993) p. 872.
\bibitem{Oshikawa-Yamanaka-Affleck-1997}
M. Oshikawa, M. Yamanaka and I. Affleck:
Phys. Rev. Lett. {\bf 78}, (1997) p. 1984.
 \bibitem{Dzyaloshinskii-1958}
I. Dzyaloshinskii: Journal of Physics and Chemistry of Solids {\bf 4}, (1958) p.241.
 \bibitem{Moriya-1960}
T. Moriya: Physical Review {\bf 120}, (1960) p.91.
 \bibitem{Uimin-Lai-Sutherland}
G. Uimin: JETP Lett. {\bf 12} (1970) p.225;
C. K. Lai: J. Math. Phys. {\bf 15} (1974) p.1675;
B. Sutherland: Phys. Rev. {\bf B 12} (1975) p.3795.
\bibitem{Takhtajan-Babujian}
L. A. Takhtajan: Phys. Lett. {\bf 87 A} (1982) p. 479; 
	H. M. Babujian: Nucl. Phys.  {\bf 215 B [FS7]} (1983) p. 317.
 \bibitem{Hakobyan-2003}
	 T. Hakobyan: J. Phys. A {\bf 36} (2003) L599
 \bibitem{Nomura-Kitazawa-1998}
	 K. Nomura and A. Kitazawa:
	 J. Phys. A {\bf 31}, (1998) p.7341.
 \bibitem{Oshikawa-2000}
	 M. Oshikawa:
	 Phys. Rev. Lett., {\bf 84}, (2000), p.1535.
 \bibitem{Yamanaka-Oshikawa-Affleck-1997}
	 M. Yamanaka, M. Oshikawa and I. Affleck:
	 Phys. Rev. Lett. {\bf 79}  (1997), p.1110
 \bibitem{Gagliardini-Hass-Rice-1998}
	 P. Gagliardini, S. Hass and T. M. Rice:
	 Phys. Rev. B {\bf 58} (1998),  p.9603. 
\end{thebibliography}
\end{document}